\documentclass{article}

\usepackage{arxiv}
\usepackage[utf8]{inputenc}

\author{
  Katharina Brauns, Christoph Scholz, Andr\'{e} Baier and Dominik Jost\\
	R\&D Division Energy Economics and Grid Operation \\
	Fraunhofer IEE\\
	34119 Kassel, Germany\\
	katharina.brauns@iee.fraunhofer.de, christoph.scholz@iee.fraunhofer.de
  \\ \\
}
\usepackage{graphicx}
\usepackage{multirow}
\usepackage{graphicx}
\usepackage{hyperref}
\usepackage[font=small, skip=0pt]{caption}
\graphicspath{ {./images/} }
\begin{document}
\title{Vertical Power Flow Forecast with LSTMs Using Regular Training Update Strategies }
%
%\titlerunning{Abbreviated paper title}
% If the paper title is too long for the running head, you can set
% an abbreviated paper title here
%

%\author{\IEEEauthorblockN{Katharina Brauns, %Christoph Scholz, Andr\'{e} Baier and Dominik %Jost\\
%	R\&D Division Energy Economics and Grid Operation \\
%	Fraunhofer IEE\\
%	34119 Kassel, Germany\\
%	katharina.brauns@iee.fraunhofer.de, %christoph.scholz@iee.fraunhofer.de}}

\maketitle              % typeset the header of the contribution
\newcommand{\rf}{\textit{RF}}
\newcommand{\RF}{\textit{Random Forests}}
\newcommand{\xgb}{\textit{XGB}}
\newcommand{\XGB}{\textit{XG-Boost}}
\newcommand{\et}{\textit{ET}}
\newcommand{\elanet}{\textit{e-net}}
\newcommand{\ELANET}{\textit{Elastic-Net}}
\newcommand{\rmse}{\footnotesize{\textit{RMSE}}}
\newcommand{\mae}{\footnotesize{\textit{MAE}}}
\newcommand{\base}{\footnotesize{\textit{Baseline}}}
\newcommand{\lstm}{\footnotesize{\textit{LSTM}}}
\newcommand{\lstms}{\footnotesize{\textit{LSTMs}}}

\begin{abstract}
The strong growth of renewable energy sources and the high volatility in power generation of these sources, as well as the increasing amount of volatile energy consumption is leading to major challenges in the electrical grid. In order to ensure safety and reliability in the electricity grid, the power flow in the grid needs to be observed to prevent overloading. Furthermore, the energy supply and consumption need to be continuously balanced to ensure the security of energy supply. Therefore a high quality of power flow forecasts for the next few hours within the grid are needed. In this paper we investigate forecasts of the vertical power flow at transformer between the medium and high voltage grid. Forecasting the vertical power flow is challenging due to constantly changing characteristics of the power flow at the transformer. This is mainly a result of dynamic grid topologies, changes in the installed assets, maintenance of the transformer itself as well as the volatile generation. In this paper we present a novel approach to deal with these challenges. For the multi step time series forecasts a \textit{Long-Short Term Memory (LSTM)} is used. In our presented approach an update process where the model is retrained regularly is investigated and compared to baseline models. The model is retrained as soon as a sufficient amount of new measurements are available. This retraining should capture changes in the characteristic of the transformer that the model has not yet seen in the past and therefore cannot be predicted by the model. For the regular update process we investigate different strategies where especially the number of used epochs are considered, but also different learning rates are used. We show that our new approach significantly outperforms the investigated baseline approaches.

\end{abstract}

\section{Introduction}
\paragraph{Problem Statement}
The massive expansion of renewable energies in recent years has led to major changes in the power flows in the electrical grid. In the past, power flows were much more predictable, both on the generation side due to power plant deployment planning, as well as on the consumption side due to established standard load profiles. But now electricity flows are much more difficult to predict, largely because of the high volatility of wind and photovoltaic power generation. Furthermore, in the past a large part of electricity generation took place at the extra-high voltage grid. A fundamental paradigm shift can now be observed. The main reason is the immense generation due to the renewable energies at all grid levels, but especially at the grid levels below the 380 kV grid. In addition, consumption behavior can no longer be predicted using standard load profiles due to the increasing use of storage facilities and e-mobility. All these factors lead to unpredictable power flows in the electricity grid, especially at the transformers between the voltage levels, the so-called vertical power flows. For a safe and reliable operation of electrical grids the balance of generation and consumption always needs to be intact. Additionally, this helps to prevent overloading. Therefore it is important to know the exact state and the electrical power flows in the grid, both the current as well as the future power flow. This knowledge allows a reliable exchange of exact system states and power flows which are required by every grid operator in order to be able to optimize and initiate safety-relevant interventions. 

For these grid operation strategies a forecast of the vertical power flow for the next few hours is needed. Forecasting the vertical power flow is challenging, due to volatile generation, dynamic grid topologies, changes in the installed assets and singular events of transformers themselves like maintenance or exchanges. These influences in the grid can cause extreme changes in the characteristics of the common behavior of a transformer and are therefore very difficult to predict with the approaches in use.  

\paragraph{Solution}
In this paper we present a novel approach to deal with the described challenges when the transformer changes its characteristic drastically. First, we introduce and analyse a \textit{Long-Short Term Memory (LSTM)} \cite{lstms1997} model architecture to calculate multi step time series forecasts of a vertical power flow. A \textit{LSTM} is a special type of (recurrent) neural network \cite{rnn1988} that is typically used for time-series prediction. In order to answer the question if the provided \textit{LSTM} model architecture is suited for a vertical power flow forecast, we  compare the forecast quality to state-of-the-art baseline models. As baseline models we use two persistence models. As an addition to the first trained \textit{LSTM} architecture we introduce and investigate a new approach using a regular update of the trained model on a daily basis. As a result, we show that this process can significantly improve the forecast quality. The model is retrained as soon as a sufficient amount of new measurements are available. This update process should capture significant changes of the transformer characteristics that have not occurred in the past. As a starting point the presented results are analysed and compared from sevenhttps://de.overleaf.com/project/5e6b5f055bf39e00010183f6 different transformer. For the update strategies we investigate several solutions to take into account the influences of different hyper-parameter like e.g. the learning rate, but also parameters like the number of epochs which give the new input measurements a higher importance.  

\paragraph{Outline:} In general, our contribution can be summarized as follows:
\begin{enumerate}
	\item To the best of our knowledge we present and analyze the first deep learning approach for building a multi step time series forecast for a vertical power flow using a \textit{LSTM} model architecture.
	\item We compare our model at different forecast horizons to baseline models.  
	\item In the context of vertical power flow forecasts, we introduce and evaluate a novel regular update approach of the \textit{LSTM} model architecture, handling the challenges of forecasting the power flow at a transformer, due to changing characteristics. We show that our model significantly outperforms the baseline models in all investigated forecast horizons.
	\item We investigate different strategies for the regular update process.
\end{enumerate}

\section{Related Work}
The volatile and distributed generation of renewable energies often results in a high load on the critical electricity grid infrastructure. This poses particular challenges for transmission and distribution system operators in terms of planning their grids reinforcements and daily grid operation. For the identification of bottlenecks, grid simulations are performed, whereby load flows of the underlying grids are of major importance. Therefor, especially in recent years the research on vertical power flow forecasts became more and more important. 

\subsection{Vertical Power Flow Forecast}
However, to date, not many scientific papers have been published in this field of research. In~\cite{jost2019} the authors presented an approach for the prediction of the vertical power flow, where they used an Extreme-Learning-Machine~\cite{extremelearningmachines} as machine learning model. They also presented a post-processing technique that ''corrects'' the forecast based on the measurements of the last days. With this post-processing the authors consider recent changes of the power flow due to for example maintenance, grid expansion or switches of the grid topology. They showed that their approach is able to forecast the vertical power flow with an accurate accuracy. Furthermore they showed that the presented post-processing step further improves the forecast quality. In~\cite{wessel2019}~the authors describe an forecast concept in which the vertical power flow is first broken down into the components wind and photovoltaics feed-in and residual load. Then for each component an individual forecast is performed and the interaction between the component is considered. However, in the current stage the paper describes the concept of a decomposed vertical power flow forecast and have not been evaluated. The forecast of wind and photovoltaics power generation is also a related topic. We refer here to the work of \cite{windreview} and \cite{pvreview}.

\subsection{Multi-Step Time Series Forecasting with LSTMs}
For the prediction of the vertical power flow we use a \textit{Long Short-Term Memory (LSTM)}. The \textit{LSTM} was introduced by Hochreiter and Schmidhuber in \cite{lstms1997}. A \textit{LSTM} is an special type of a \textit{Recurrent Neural Network (RNN)}~\cite{rnn1988} with the advantage that long-term dependencies can be considered. Furthermore it is able to deal with the vanishing and exploding gradient problem. In~\cite{marino2016} Marino et al. compared \textit{LSTMs} with a \textit{LSTM}-based Sequence to Sequence (S2S) architecture for energy load forecasting. The authors showed that the \textit{S2S-LSTM} performs better than the ''normal'' \textit{LSTM}. For a detailed overview concerning multi-step forecasting strategies we refer to~\cite{BENTAIEB20127067}.

\section{Dataset for Vertical Power Flow Forecasting}\label{label:dataset}
In this paper we investigate the different forecast approaches using seven transformers. These seven transformers were selected on the basis of a first evaluation of the forecast calculated in the project EU-SysFlex\footnote{\url{https://eu-sysflex.com/}}. In sum 585 models using the \textit{LSTM} model architecture described in Figure \ref{fig:lstm_architecture} were developed and evaluated. The results were analyzed using error measures such as root mean square error (RMSE), Pearson correlation and the time series plot itself. It turned out that the used model did not perform very well for several transformers. 
The reason is due to the already explained challenges of the changing characteristics of a transformer which is shown in the next subsection (\ref{label:vert_power_flow}). 
In order to get a good starting point for evaluating the new update approach and to have a more reliable result by using seven transformer instead of just one, the transformer were chosen by the above mentioned error measures with a RMSE higher than 38\% and a Pearson correlation between 0.6 and 0.8. Additional to these error measure selection, the time series plots were analysed and the results were scaled down to seven transformers, since here the forecast were especially influenced and did not perform well due the mentioned changing characteristics of the transformer. After these analysis 7 transformers were found to be a good starting point for the analysis of the regular update process.

\subsection{Vertical Power Flow Measurements}\label{label:vert_power_flow}
\begin{figure}[!h]
\centering
\includegraphics[scale=0.25]{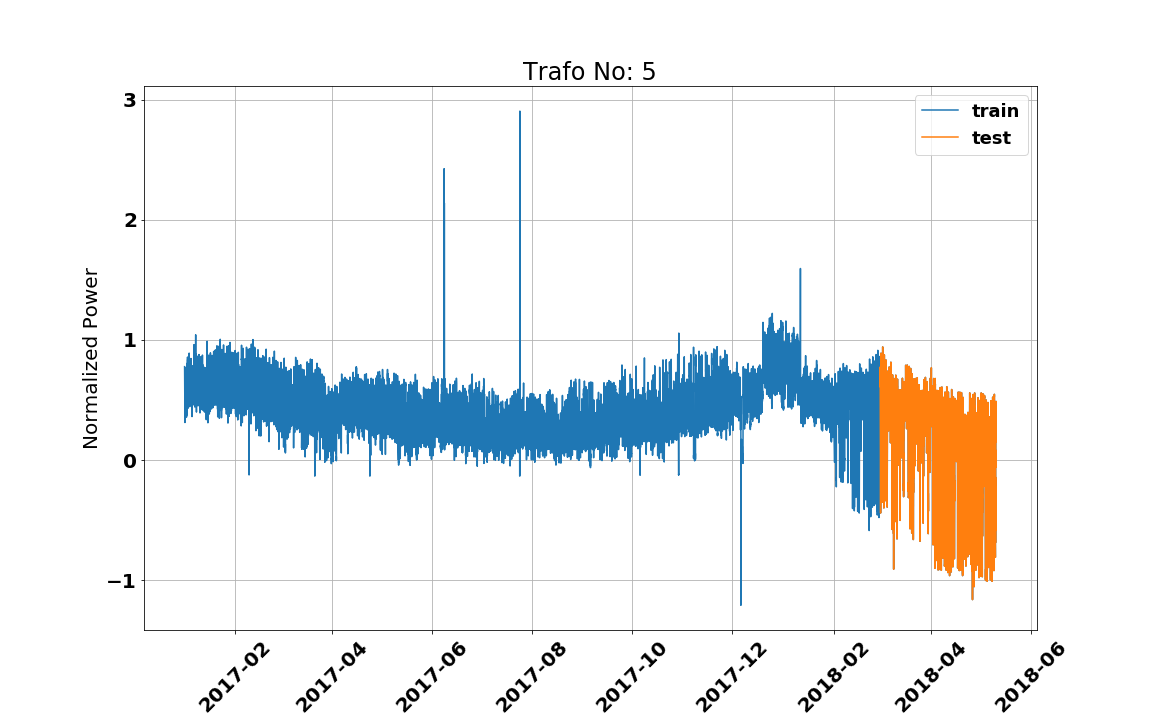}
\caption{Time series of transformer $5$ with training data in blue and test data in orange. Especially the changed characteristic of the transformer in the test data can be seen.}
\label{fig:ts_example_characteristics}
\end{figure}

As mentioned before, the vertical power flow is the power flow in the electrical grid between different voltage levels. In the project EU-SysFlex mainly the vertical power flow between the medium to the high voltage grid is used. This voltage grid is influenced by a dynamic grid topologie, which can result from changes in the switching state, new installed assets or maintenance work on the transformer itself. New assets could be a completely new installed wind farm or photovoltaic system which could result in a high amount of additional generation in the power flow of the transformer. On the other hand it could also be a newly connected industry or commercial enterprise which adds up to a higher consumption at the transformer. 
This results in changing characteristics in the transformer behaviour, which makes it difficult to predict. In the past the consumption was easy to predict since it had a periodic behaviour where normal standard load profiles could be used to forecast the consumption. But due the rising share of a volatile consumption through the usage of storages and e-mobility in the future, this is getting more and more complicated to forecast. The share of volatile generation, especially from wind farms and photovoltaic systems, is currently rising continually, which also makes the forecast more complicated. Some transformers are only connected to consumers and some only to wind or photovoltaic farms. Depending whether the consumption or the generation is higher, the values of the vertical power flow can be positive or negative.  
In Figure \ref{fig:ts_example_characteristics} an example of one transformer with a changing characteristic is shown. The blue line is the data of the training data set and the orange line indicates the test data set. To get a better overview in the figure, the training data from the year 2016 was cut out, which has a very similar behaviour to the year 2017. It can be seen that the training data has a different scaling than the test data and that from around the February 27, 2018 the values are getting more negative which means that high generation is added to the transformer. In this case the original trained model has no knowledge of the additional installed power and cannot represent this changing in the characteristic. 
The vertical power flow which is used in this paper has a time resolution of 15 minutes and will be updated through an online process every 15 minutes as well. From the available meta data only geographical information are used for combining the transformer location to the numerical weather forecasts.

\subsection{Weather data}
Next to the vertical power flow measurements, the numerical weather forecast are used as input for the model. The weather forecast are delivered from the IFS from European Centre for Medium-Range Weather Forecasts. The IFS has a high forecast quality, but a rather low spatial resolution of 9km and a time resolution of three hours for the online data. The forecast horizon is from 0 to 10 days, which makes it flexible in the usage once the forecast horizon needs to be extended which is the intention in the project EU-SysFlex. Additionally, the forecasts are updated two times a day, in the morning at around 9am and in the evening at around 6pm. It has a wide range of parameters where we are using a set of the following parameters\footnote{All parameters in italic can be found with a more detailed description at\\ \url{https://www.ecmwf.int/en/forecasts/datasets/set-i\#I-ii}}:
\begin{itemize}
 \item Sun position information\footnote{Calculated with the Pysolar package \url{https://pypi.org/project/Pysolar/}}: Altitude, azimuth, solar radiation
 \item Directions of the wind speed in 10m and 100m (\textit{10u}, \textit{10v} and \textit{100u}, \textit{100v})
 \item Temperature in 2m (\textit{2t}) and dewpoint temperature in 2m (\textit{2d})
 \item Forecast albedo (\textit{fal}) and surface solar radiation downwards (\textit{ssrd}) 
 \item Surface pressure (\textit{sp}) and total precipitation (\textit{tp})
 
\end{itemize}

\section{Forecast Models}
\subsection{Short-Term Vertical Power Flow Forecast Using LSTM Models}\label{label:short-term fc}
\begin{figure}[t]
\centering
\includegraphics[width=0.8\textwidth]{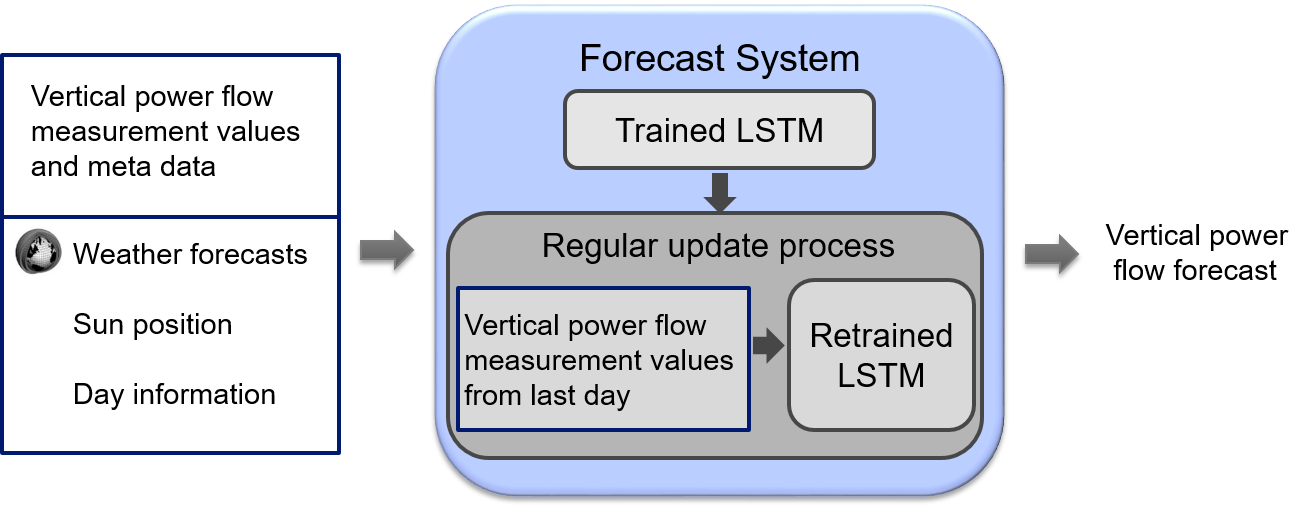}
\caption{Forecast system of the vertical power flow forecast (right side) with input data (left side) and used machine learning algorithm (middle) in the way how it is used in the project EU-SysFlex and with addition of the regular update process (middle lower part), which is proposed here in the paper.}
\label{fig:forecast_system}
%\vspace*{-6mm}
\end{figure}

Within the scope of the project EU-SysFlex, a demonstrator for a distribution grid in Germany will be built. This demonstrator has the goal to create optimized active and reactive power schedules for an optimized congestion management in order to support the DSO to provide flexibility of the distribution system and voltage control as an ancillary service to the TSO. In order to be able to create the required schedules of active and reactive power for the next hours and days, forecasts with a high quality of the vertical power flows at the transformers of the medium and high voltage level are necessary.
A vertical power flow is defined as power flow between electrical grids with different voltage levels, which can take both negative and positive values, depending on whether consumption or production predominates. 
The forecast for the vertical power flow is calculated for each transformer in the medium voltage connected to the high voltage grid.
In Figure \ref{fig:forecast_system} we show the developed forecast system with its input data, its machine learning algorithm and the output as a vertical power flow forecast. For training the models the vertical power flow measurement values\footnote{The data set was provided in the project and must be treated confidentially.}, which are transferred from the DSO, are used including a status information about the reliability of the measurement value (0 for true, 1 for false). Additionally, a numerical weather forecast, the sun position, day information data with the hour of the day and the weekday is used as input data for training the models. The vertical power flow input values have a time resolution of 15 minutes for both historical and online data. The forecast horizon of the calculated vertical power flow forecast is 48 hours. The forecast is also generated with a time resolution of 15 minutes and is regularly generated every 15 minutes and delivered to the DSO.
In this paper we use the same constraints for the multi step time series forecast for the time resolution, the forecast horizon and the update cycle for the delivery as in the project.

\begin{figure}[ht]
\centering
\includegraphics[scale=0.33]{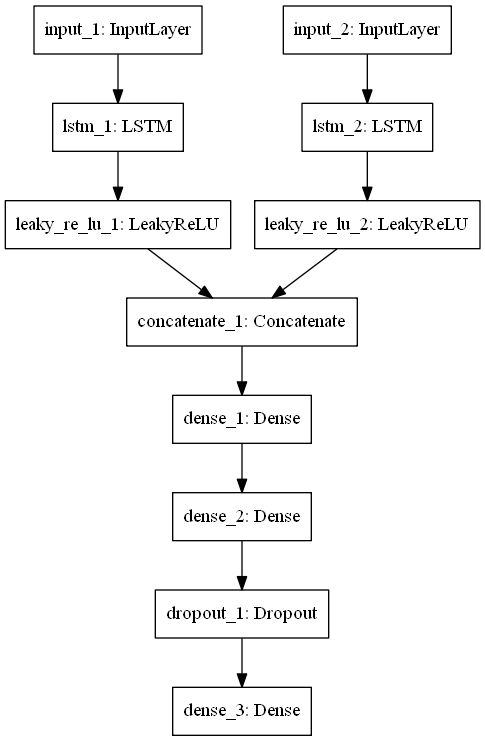}
\caption{Architecture of the \textit{LSTM} model. }
\label{fig:lstm_architecture}
%\vspace*{-5mm}
\end{figure}

As introduced, for the training of the vertical power flow forecast a deep neural network architecture with \textit{LSTMs} is used. This architecture is shown in Figure \ref{fig:lstm_architecture} and is described in the following. The advantage of using a deep neural network with several representation layers is that no time series feature engineering is necessary. "The model learns the representation layers together at the same time so that all depending features are adjusted and all modifications to the model contribute to the same goal"\cite{Chollet.2018}. "With more layers a more complex representation of the data can be achieved"\cite{Chollet.2018}. Unfortunately, there is no framework up to the time this paper was written which can automatically find an adequate architecture for multi-step time series regression problems.
%with e.g. the best amount of layers to use for time series regression analysis.
For the work of this paper we use 4 layers. %The whole model architecture is described in the following.
For each transformer a \textit{LSTM} model architecture is generated and used. The weights created in the training, which are saved separately for each transformer, are then added to the \textit{LSTM} model to calculate the forecast for each transformer.
The first layers of the model architecture are two parallel input layers which describes the true measurement values of the vertical power flow (\texttt{input\_1}) and in (\texttt{input\_2}) all the other features, the numerical weather forecasts, the sun position and the day information. These input layers are followed by two \textit{LSTMs} that use the input time series to calculate the hidden output states separately and then concatenate them into a next layer. As activation function for the \textit{LSTMs} the Leaky Rectified Linear Unit (LeakyReLU) is used. After the concatenated layer follow two Dense layer. An important component in deep neural networks is the regularisation for the avoidance of over-fitting. Thus, the two Dense layers are extended by a Dropout layer and the recurrent dropout is used in the \textit{LSTMs}. The last layer is a fully connected output layer. As optimizer we choose Adam with the two different learning rates of 0.001 and 0.01 and the loss function with the mean absolute error for the first trained models and mean squared error for the regularly updated models. For all densely connected layers we applied Rectified Linear Unit (ReLu) as activation function. The additional hyper-parameters are listed in the Table \ref{tbl:1} in section \ref{label:exp_setting_1}. We here note that this subsection (\ref{label:short-term fc}) is an adapted version of section 2 within the deliverable of EU-SysFlex\footnote{D6.2 'Forecast: Data, Methods and Processing. A common description'}.  

\subsection{Regular Training Update Strategies Using LSTM Models}\label{label:regular_updates}
As mentioned in the description of the data set, especially the changing of the characteristics of a transformer makes it difficult to perform high quality forecasts. This is indicated in Figure \ref{fig:ts_example_characteristics} where the test data changes the behaviour of the vertical power flow enormous. The idea is to handle exact these kind of changes with a regular update respectively retraining of the model. In this paper we use a retraining on a daily basis. A description of the forecast system using the regular update process is already shown in Figure \ref{fig:forecast_system}. Since the vertical power flow measurements are delivered in an online process every 15 minutes the data from the last day is used as input to the previous trained \textit{LSTM} model at the first step and to the retrained model afterwards. The model architecture described in Figure \ref{fig:lstm_architecture} is not changed, but hyper-parameters like the learning rate and the number of epochs, see Table \ref{tbl:updates} in section \ref{label:exp_setting_updates}. The main reason for choosing these parameters is to weigh the current available data higher then the historical data which was already used for training the first model.        

\subsection{Baseline Models}
Persistence models are usually very simple models, only using the time series itself as input for the forecasting. Although they are very simple, they are often still hard to beat, especially for low forecast horizon up to 1 hour. That is why we chose the following two persistence models. 

\paragraph{\textbf{Persistence Last Measurement:}}\label{label:pers_last_meas}
The simplest way of using a persistence model, is by use only the last available measurement value and repeat this for all forecast horizon. At the time the forecast is calculated, this means that for the next 48 hours (maximum used forecasts horizon) the predicted values have all this same value of the last available measurement.  

\paragraph{\textbf{Persistence Last Day:}}\label{label:pers_last_day}
Another way of using a persistence model for forecasting is to use the available measurements from the last day. This means by predicting the next hour value valid at e.g. 3pm at day 1, then the forecast gets the same value as the measurement had at the day before (day 0) at 3pm and so on. If the forecast horizon exceeds the 24 hours, it just starts over with the last day, meaning that for predicting 25 hours ahead, which would be again a value valid at 3pm for day 2, it again gets the same value from day 0 at 3pm. Summarized, creating this kind of persistence at the time the forecast is calculated, just means it predicts the exact same values from the day before for the first day (forecast horizon 0 to 24 hours) and repeats this by predicting the second day (forecast horizon 25 to 48 hours). 

\section{Experiments}
\subsection{Setting: Vertical Power Flow Forecast with LSTMs}\label{label:exp_setting_1}
For the development and evaluation of a forecast usually the input data needs to be prepared in a pre-processing step and also be split into training, validation and test data sets.
The pre-processing step is mainly needed for feature engineering which is important for using machine learning algorithm. As explained in \ref{label:short-term fc} for a deep learning approach it is not required and actually already included in the deep neural network architecture. Thus, the assumption is that we do not have to e.g. pre evaluate which feature of the input data is important and which should be dropped, since the model itself decides which to use. Since we also have the status information of the vertical power flow measurements, i.e. a status of 1 means unreliable measured value and 0 means reliable measured value, we do not perform any real data cleansing beforehand. For the forecast using the \textit{LSTM} model without regular updates historical data from January 2016 to May 2018 is available for all transformers. The splitting of the historical data into training, validation and test data takes place for exact dates using the 1st of January 2018, the 1st of March 2018 and till the end of the available data with the focus on getting the most possible training data set.
For the combination of the transformer with its corresponding numerical weather forecast, coordinates of both locations are mapped. 
The model hyper-parameters are listed in the following Table \ref{tbl:1}. The selection of these hyper-parameters was based on \cite{Greff.2017} and on the basis of already gained experience. %using the example of some experimental applications connected to other projects.
For the verification, the results were compared with a hyper-parameter optimization using an Automated Machine Learning (AutoML) approach investigated by Salz in \cite{Salz.2020}. It could be shown that the used hyper-parameters from Table \ref{tbl:1} are leading to similar results than the results from the AutoML approach. %
{
    \renewcommand{\arraystretch}{1.01}
\begin{table}[ht]
\centering
% To place a caption above a table
\begin{tabular}{|c|c|} \hline
  \textbf{Model Hyper-Parameter} & \textbf{Number} \\\hline
 Number neurons LSTM layer & 100  \\\hline
 Number neurons dense layer 1 & 500 \\\hline
 Number neurons dense layer 2  & 500 \\\hline
 Recurrent dropout & 0.5 \\\hline
 Dropout  & 0.5 \\\hline
 Batch size  & 192 \\\hline
 Number epochs  &  40 \\\hline
 Steps per epoch  & 50 \\\hline
 learning rate & 0.001 \\\hline
 loss function & MAE \\\hline
\end{tabular}
% Or to place a caption below a table
\caption{Listing of the hyper-parameters and their values used for the \textit{LSTM} model.}
\label{tbl:1}
\end{table}%
%\vspace*{-10mm}
}
\subsection{Setting: Vertical Power Flow Forecast with LSTMs Using Regular Update Strategies} \label{label:exp_setting_updates}
In section \ref{label:regular_updates} the regular update process is described. Using the validation and test data set, we retrain the model each day using the input data from the previous day including 96 time steps. The validation data is needed, since the previous trained model has not seen this data yet. It was only used for validating the training. For the evaluation, we only use the test data which is saved after each update cycle. In Table \ref{tbl:updates} only the changed hyper-parameters (compared to Table \ref{tbl:1}) are listed. Since we investigate different update strategies more than one value is given for the number of epochs and learning rate. We evaluate  regular updates using the following number of epochs: 5, 10, 15 and 20 and a learning rate of 0.01 and 0.001.

{
    \renewcommand{\arraystretch}{1.02}
\begin{table}[ht]
\centering
% To place a caption above a table
\begin{tabular}{|c| c|} \hline
  \textbf{Model Hyper-Parameter} & \textbf{Number} \\\hline
 Number epochs  &  5, 10, 15, 20  \\\hline
 Steps per epoch  & 1 \\\hline
 learning rate & 0.01, 0.001 \\\hline
 loss function & MSE \\\hline
\end{tabular}
% Or to place a caption below a table

\caption{Listing of the hyper-parameter and their values used for the \textit{LSTM} model with regular updates}
\label{tbl:updates}
%\vspace*{-13mm}
\end{table}%
}

\subsection{Evaluation Procedure and Error Measures}
For the evaluation of the forecast results two different normalization methods were used. The first one is used for normalizing the input data of the model to similar scales which is in general done for training a neural network. For this normalization the data is subtracted by its mean and then divided by its standard deviation, both calculated from the training data set, i.e.

\begin{equation} \label{equat_norm_1}
x_{scaled} = \frac{x - \bar{x} }{\sigma},                                                        \end{equation}
where x is the original value, \(\bar{x}\) the mean and \(\sigma\) is the standard deviation. To fully pre-process the input data for the calculation of the deep neural network models the data is allocated into batches. 

For the training procedure the validation data set is evaluated through 40 epochs. The usage of the 'EarlyStopping' may prevent the passing through all 40 epochs. The best weights are saved and used together with the test data, which are processed similarly to the training data, for evaluation of the models. 
In order to calculate error measures like the RMSE and the Pearson correlation, which we use mainly to evaluate the forecast results, a normalization of the forecast time series values between 0 and 1 is necessary. Therefor, these values need to be inverted back to absolute values and then to be re-normalized with the following equation: 

\begin{equation} \label{equat_norm_2}
x_{scaled} = \frac{x - Q_{0.03}(x)} {Q_{99.7}(x) - Q_{0.03}(x) },                               \end{equation}

where x is the original value and \(x_{scaled}\) is the normalized value. For the prevention of using outliers for the maximum and minimum values of x, the quantile values of the 99.7\% quantile (\(Q_{99.7}\)) and the 0.03\% quantile (\(Q_{0.03}\)) respectively are used instead as maximum and minimum.

\section{Results}
In this section the results for all models and all update strategies described in the previous chapters are compared and analyzed. In the following the results are mainly aggregated for a better overview, but to discuss the results in more detail we chose only one transformer exemplary.

\subsection{Results of Regular Training Update Using Different Strategies} \label{label:result_diff_strategies}
\begin{figure}[!h]
\centering
\includegraphics[width=0.75\textwidth]{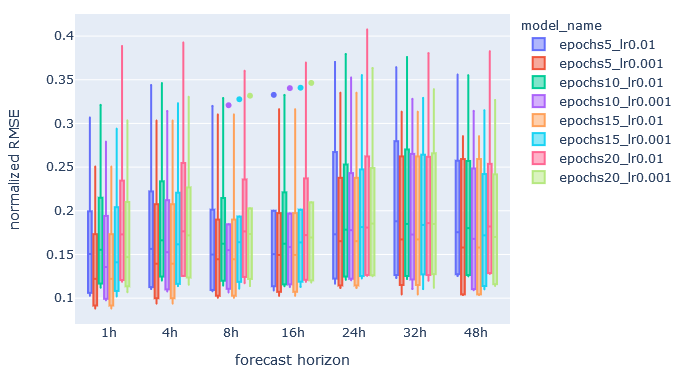}
\caption{Boxplot of the normalized RMSE with a comparison of the different update strategies using epochs 5, 10, 15 and 20 and a learning rate of either 0.01 or 0.001.}
\label{fig:boxplot_strategies}
%\vspace*{-8mm}
\end{figure}

\begin{figure}[!htb]
a){
\begin{minipage}{1\columnwidth}
\centering
\includegraphics[width=9.5cm]{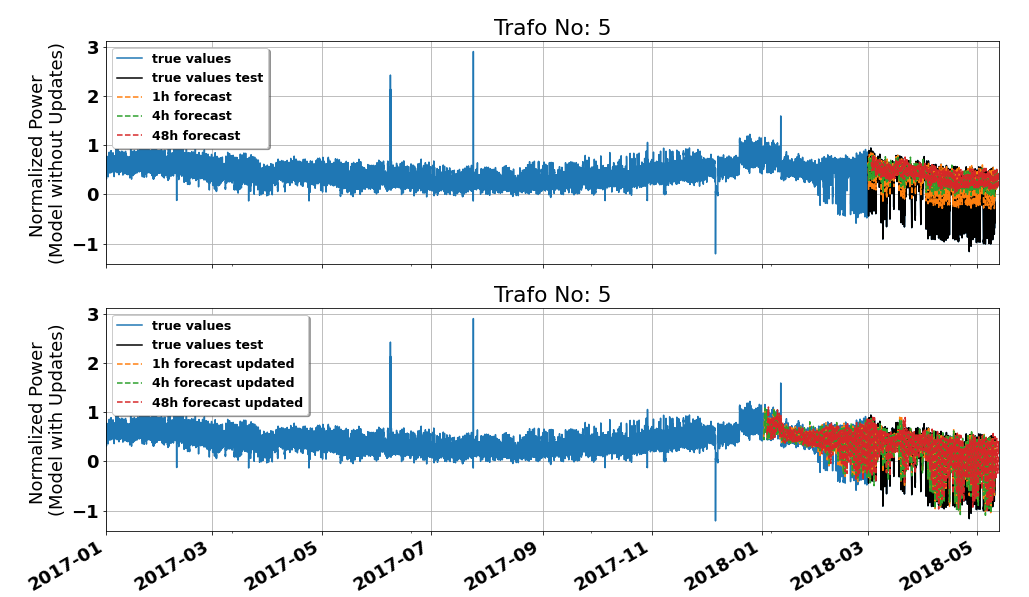}
\end{minipage}
%\label{fig:ts_plot_without_with_updates_a}
}
%\hfill
b){
\begin{minipage}{1\columnwidth}
\centering
\includegraphics[width=9.8cm]{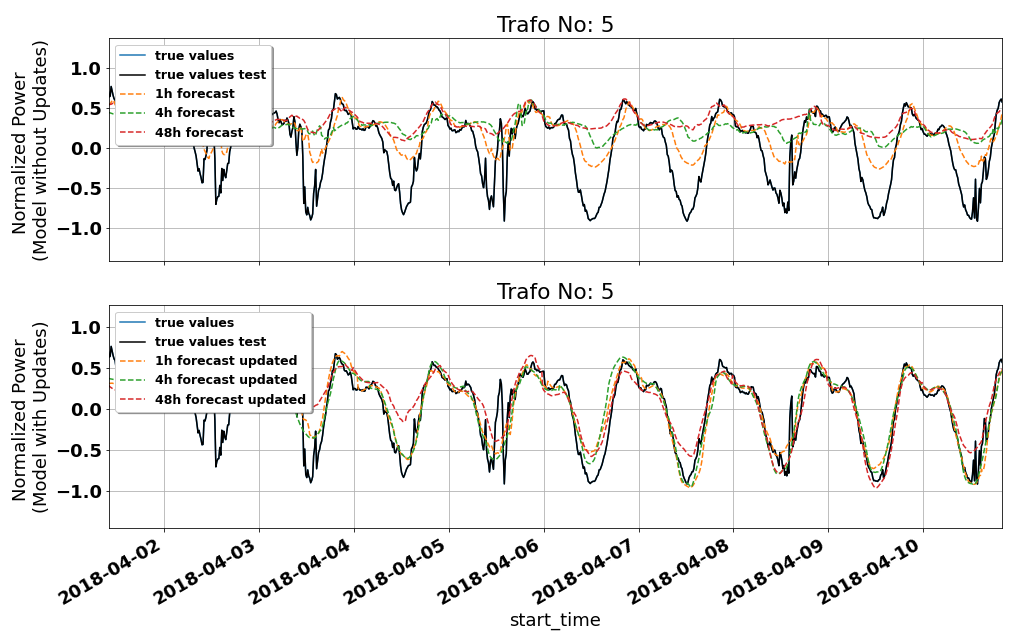}
\end{minipage}
%\label{fig:ts_plot_without_with_updates_b}
}
\caption{a) Time series plots with the forecasts of the 1 hour, 4 hour and 48 hour forecast horizon and true measurement values. The blue line marks the training and validation data and the black line the test data. The upper plot contains the \textit{LSTM} model results without and the lower plot the \textit{LSTM} model with a regular retraining of the model.
b) This time we zoomed in and use a time span of about 10 days.}
\label{fig:ts_plot_without_with_updates_a}
%\vspace*{-5mm}
\end{figure}

The main problem we like to solve by using regular training updates is to improve the vertical power flow forecast which is negative impaired by the changing characteristics of transformers described in \ref{label:vert_power_flow}. As shown in Figure \ref{fig:ts_example_characteristics}, the behaviour in the test data has changed extremely, so that the previously trained model can have no knowledge of it and therefore performs poorly. With adding a daily re-training of the existing model the new behaviour of the data should be learned and then be correctly represented by the model. For the daily training we investigate different strategies like using a different number of epochs and learning rate in order to weight the new available input data higher then the historical data used for the first training. In Figure \ref{fig:boxplot_strategies} the results for the different update strategies are summarized. We have eight models where the different number of epochs are combined either with a learning rate of 0.01 or 0.001. Each strategy is evaluated for 7 different forecast horizons which are represented by the x-axis. As error measure the normalized RMSE is used and represented by the y-axis. 
%\begin{figure}[!h]
%\centering
%\includegraphics[width=0.80\textwidth]{images/boxplot_comparison_all_update_strat%egies_rmse.png}
%\caption{Boxplot of the normalized RMSE with a comparison of the different update %strategies using epochs 5, 10, 15 and 20 and a learning rate of either 0.01 or %0.001.}
%\label{fig:boxplot_strategies}
%\vspace*{-8mm}
%\end{figure}
%It is shown that in principle a higher number of epochs gives a better performance, whereby the best performance is not necessarily achieved at 20 epochs, but the results are similarly good at 15 epochs. A learning rate of 0.01 results in most cases in a better performance. For further discussing the results, we chose the update strategy with an epoch number of 15 and a learning rate of 0.01 as the best performing one.
It is shown that in principle a higher number of epochs does not automatically gives a better performance. In contrary, the best performance is achieved at 5 epochs. A learning rate of 0.001 results in most cases in a better performance. For further discussing the results, we chose the update strategy with an epoch number of 5 and a learning rate of 0.001\footnote{For future work it still needs to be evaluated whether the number of used epochs and the learning rate provide high  quality forecasts on more test periods. For now, our intention is to show that the suggested approach does improve the forecasts.}. %as the best performing one.
In Figure \ref{fig:ts_plot_without_with_updates_a} we compare the forecasts without and with a regular retraining for the transformer with the number 5. Since the forecasts calculated in the regular update process need to start with the validation data, it is also plotted in the lower plot (\ref{fig:ts_plot_without_with_updates_a}a). The blue time series mark the training/validation data and the black time series mark the test data from the vertical power flow true measurements. In both plots the forecasts are plotted for three forecast horizon, the 1 hour, 4 hour and 48 hour horizon. In the upper plot (\ref{fig:ts_plot_without_with_updates_a}a and \ref{fig:ts_plot_without_with_updates_a}b) is the forecast without any update of the \textit{LSTM} model shown. We see that (on the test data) the predicted values are too low for each forecast horizon.. In contrast, the forecast using the regular update process reaches the entire range of newly scaled test data which can be seen in the lower plot (\ref{fig:ts_plot_without_with_updates_a}a and \ref{fig:ts_plot_without_with_updates_a}b).%
%\vspace*{-4mm}
%\begin{figure}[!htb]
%a){
%\begin{minipage}{1\columnwidth}
%\centering
%\includegraphics[width=9.8cm]{images/486542070_comparison_with_and_without_updat%#s_ts_subplots_copy.png}
%\end{minipage}
%%\label{fig:ts_plot_without_with_updates_a}
%}
%%\hfill
%b){
%\begin{minipage}{1\columnwidth}
%\centering
%\includegraphics[width=9.8cm]{images/486542070_comparison_with_and_without_updat%#s_ts_subplots_zoom_2_copy.png}
%\end{minipage}
%%\label{fig:ts_plot_without_with_updates_b}
%}
%\caption{a) Time series plots with the forecasts of the 1 hour, 4 hour and 48 %hour forecast horizon and true measurement values. The blue line marks the %training and validation data and the black line the test data. The upper plot %contains the \textit{LSTM} model results without  and the lower plot the %\textit{LSTM} model with a regular retraining of the model.
%b) This time we zoomed in and use a time span of about 10 days.}
%\label{fig:ts_plot_without_with_updates_a}
%\vspace*{-5mm}
%\end{figure}
In order to get a more complete, somewhat deeper insight, also with regard to the forecast horizons, we have zoomed into the figure once again. This can be seen in Figure \ref{fig:ts_plot_without_with_updates_a}b. The correlation of the forecasts, meaning with all shown forecast horizon, with a regular update process are much higher than without it. 
%\begin{figure}[h]
%\centering
%\includegraphics[width=\textwidth]{images/4865542070_comparison_with_and_without_updates_ts_s%ubplots_zoom_2.png}
%\caption{Time series plots with the forecasts %of the 1 hour, the 4 hour and the 48 hour forecast horizon and true measurement values. The black line marks the test data. The upper subplot contains the \textit{LSTM} model results without and the lower subplot the \textit{LSTM} model with a regular retraining of the model. This time we zoomed in and use a time span of about 10 days.}
%\label{fig:ts_plot_without_with_updates_zoom}
%\end{figure}
%\begin{figure}[h]
%\centering
%\includegraphics[width=12.1cm]{images/486542070_scatter_plot_without_and_with_upd%ates_epochs15_lr0_01_4_fc_hor.png}
%\caption{Scatter plots for 4 different forecast horizon (1h, 4h, 16h, 48h) for %transformer 5. The true measurement values are plotted against the forecast %values. In the upper plot row, the forecast results from the \textit{LSTM} model %without using the regular retraining and in the lower plot row, the forecast %results from the \textit{LSTM} model with regular retraining is shown.}
%\label{fig:scatter_without_with_updates}
%\vspace*{-6mm}
%\end{figure}
\begin{figure}[h]
\centering
\includegraphics[width=11.0cm]{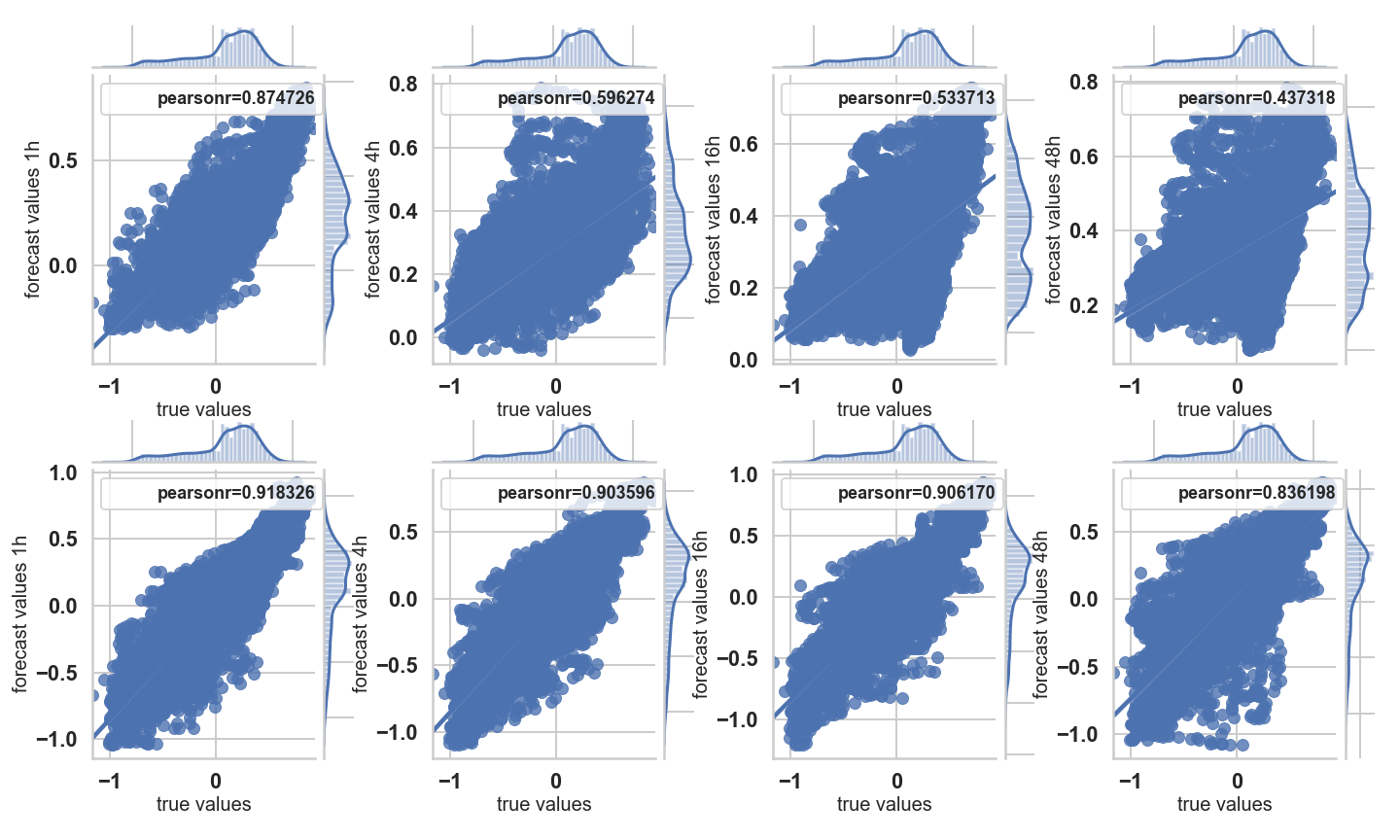}
\caption{Scatter plots for 4 different forecast horizons (1h, 4h, 16h, 48h) for transformer 5. The true measurement values are plotted against the forecast values. In the upper plot, the forecast results from the \textit{LSTM} model without using the regular retraining, and in the lower plot, the results from the \textit{LSTM} with regular retraining is shown.}
\label{fig:scatter_without_with_updates}
\vspace*{-6mm}
\end{figure}
This conclusion is also presented in Figure \ref{fig:scatter_without_with_updates} where a scatter plot for four forecast horizon and again without and with a regular retraining is used. To the previous used forecast horizon in Figure \ref{fig:ts_plot_without_with_updates_a}a and Figure \ref{fig:ts_plot_without_with_updates_a}b, the 16 hour horizon is added. The scatter show the relation between the forecasts (y-axis) and the true measurement values (x-axis). A line for comparison shows the case if the model had learned the exact relationship between the explanatory variables and the vertical power flow measurements. In the upper plot, there is a moderately strong linear relationship between the forecasts and the measurement values without updating the model. The correlation gets worse the higher the forecast horizon gets (from left to right). %In contrast to the figure below, where the regular update process is used. Here, a strong linear relationship between forecasts and measurements with a high correlation for each forecast horizon can be seen. 
In contrast to the figure below, where the regular update process is used. Here, a stronger linear relationship between forecasts and measurements with a higher correlation for each forecast horizon can be seen. %The correlation does not get worse for each forecast horizon, only for the last horizon of 48h it is worse than for the smaller horizons. 

\subsection{Results and Comparison: LSTM, LSTM with Updates and Baseline Models}
\begin{figure}[!h]
\centering
\includegraphics[width=10.0cm]{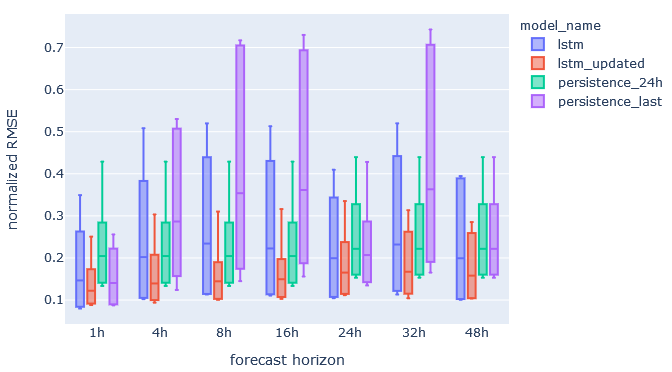}
\caption{Boxplot of normalized RMSE with a comparison of different models, first trained \textit{LSTM} without updates (lstm), \textit{LSTM} with updates (lstm\_updated), persistence using last 24 hours (persistence\_24h), persistence of the last value (persistence\_last).}
\label{fig:boxplot_models}
\vspace*{-6mm}
\end{figure}

\begin{figure}[!h]
\centering
\includegraphics[width=8.0cm]{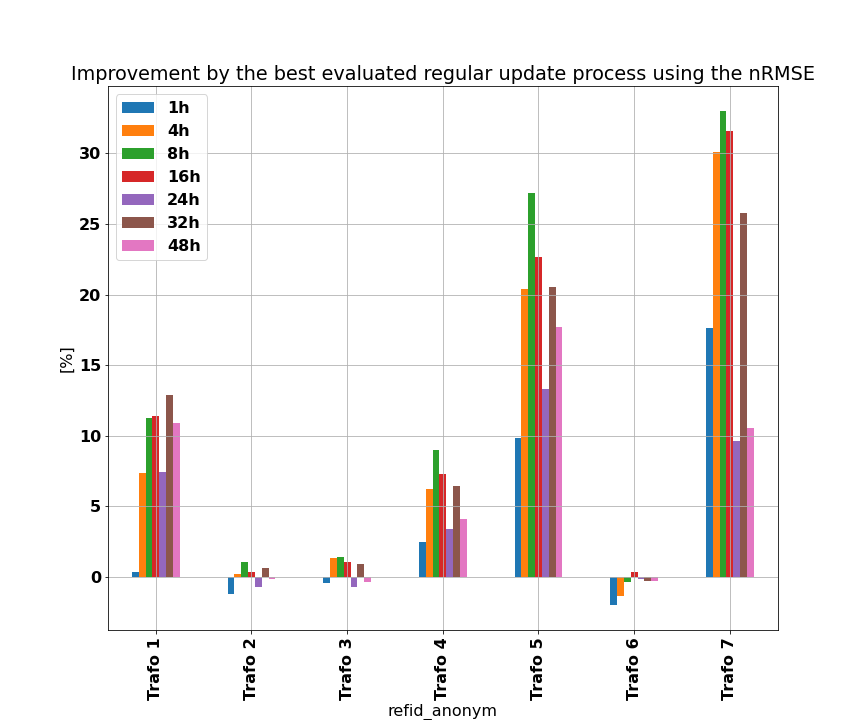}
\caption{Improvement of the normalized RMSE gained by using the regular training update compared to the \textit{LSTM} model without regular updates.}
\label{fig:fc_improvement}
%\vspace*{-4mm}
\end{figure}

In order to evaluate the performance of the used \textit{LSTM} model architecture with and without using the regular update process, we compare the forecast results with the two baseline models which is shown in Figure \ref{fig:boxplot_models}. The results are again presented aggregated over all seven transformers and for seven forecast horizon: 1h, 4h, 8h, 16h, 24h, 32h and 48h. As error measure the normalized RMSE is used. It is clear to see that the \textit{LSTM} model architecture using the regular update of the model outperforms all other models. Regarding the \textit{LSTM} model without update process, it only exceeds in the 1h forecast horizon the persistence model using the last day, described in \ref{label:pers_last_day}. For the other forecast horizon it outperforms only the persistence model using the last available measurement value, described in \ref{label:pers_last_meas}, for the forecast horizon of 4h, 8h, 16h and 32h.
Apparently the forecast horizon of 24h and 48h are more easy to predict, especially for both persistence models. The reason could be that the vertical power flow has a similar repeating pattern on a daily basis which is usually the case if the consumption is predominating over the volatile generation. This is actually for most of the seven transformers the case.
%\begin{figure}[!h]
%\centering
%\includegraphics[width=10.9cm]{images/boxplot_comparison_all_models_rmse.png}
%\caption{Boxplot of normalized RMSE with a comparison of different models, first %trained \textit{LSTM} without updates (lstm), \textit{LSTM} with updates %(lstm\_updated), persistence using last 24 hours (persistence\_24h), persistence %of the last value (persistence\_last).}
%\label{fig:boxplot_models}
%%\vspace*{-1mm}
%\end{figure}
Finally, we evaluate for all seven transformer and all seven forecast horizon the improvement of the normalized RMSE gained by using the regular training update (see Figure \ref{fig:fc_improvement}).% For all transformer an improvement can be achieved, although 3 transformers only provide an improvement slightly less than 5\%. On the other hand, the other four transformers show an improvement of more than 8\% for almost all forecast horizons except the 1h horizon. But at least a 4\% improvement can be achieved here.
~An improvement can be achieved for all transformers, although for 3 transformers this is only true for some forecast horizons. These transformer provide a maximum improvement of about 1.4\%. On the other hand, the other four transformers show a mean improvement of more than 7.5\% for all forecast horizons. However, in average a performance reduction of 1.2\% must be accepted for the other 3 transformers.
%\vspace*{-2mm}

%\begin{figure}[!h]
%\centering
%\includegraphics[width=12cm]{images/forecast_improvement_by_updates%_epochs15_lr0_01_all_trafos_rmse.png}
%\caption{Improvement of the normalized RMSE gained by using the %regular training update compared to the \textit{LSTM} model without %regular updates.}
%\label{fig:fc_improvement}
%\vspace*{-4mm}
%\end{figure}
%%\vspace*{-6mm}

\vspace*{-6mm}

\section{Discussion and Outlook} 
In this paper we present a novel approach using a regular update process in combination with a previous trained \textit{LSTM} model and compare both performances. In addition we compare our model with two persistence baseline models. In a first evaluation the optimal strategy for the update process was determined. For this purpose, the best number of epochs and the best learning rate were identified. %In a second evaluation the \textit{LSTM} model using the resulted best update strategy is used for the comparison to the other models. Overall, the results show that high performance is achieved by our new approach. It performs significantly better for all forecast horizon and all transformer than the other considered models. The performance is evaluated with the normalized RMSE and the Pearson correlation. A significant improvement of the normalized RMSE with respect to the \textit{LSTM} model without regular update process of on average 13.5\% across all horizons was achieved.
In a second evaluation the \textit{LSTM} model using the resulted best update strategy is used for the comparison to the other models. Overall, the results show that high performance is achieved by our new approach. In average it performs significantly better for all forecast horizons and all transformers than the other considered models. In our experiments a significant improvement of 8\% (in average) could be observed using the \textit{LSTM} model with regular update process.
For future work, to further improve our models,
we want to consider transfer learning approaches. This offers the possibility to cover further challenges, especially those posed by the changing characteristic of the transformers. Additionally to the presented work, we compared the \textit{LSTM} model to an Encoder Decoder architecture using \textit{ConvLSTM2D}\cite{CNNLSTM} layers and Attention layers. Unexpectedly, we did not achieve a better result, so that we did not include the results into this paper. There should be further work to investigate the Encoder Decoder architecture more deeply. Finally, we suggest to compare our results to the method described in \cite{jost2019}.% and maybe even analyse a combination of both.

%\section{Acknowledgement}
\section*{Acknowledgement}
This work is based on results from the EU-SysFlex project. This project has received funding from the European Union's Horizon 2020 research and innovation program under grant agreement No 773505. The authors are solely responsible for this publication. 

%The project EU-SysFlex has received funding from the European Union’s Horizon 2020 research and innovation program under grant agreement No 773505.
\bibliographystyle{splncs04}
\bibliography{references}

\end{document}